\documentclass[aps,prx,showpacs,twocolumn,superscriptaddress]{revtex4}
\usepackage{graphicx}
\usepackage{amssymb}
\usepackage{epsfig}
\usepackage{graphicx}
\usepackage{amsmath}
\usepackage{array,color}

\begin{document}

\title{Tunneling-assisted Spin-orbit Coupling in Bilayer
Bose-Einstein Condensates}
\author{Qing Sun}
\affiliation{Department of Physics, Capital Normal University,
Beijing 100048, China}
\author{Lin Wen}
\affiliation{College of Physics and Electronic Engineering, Chongqing Normal University, Chongqing, 401331, China}
\author{W.-M. Liu}
\affiliation{Beijing National Laboratory for Condensed Matter
Physics, Institute of Physics, Chinese Academy of Sciences, Beijing
100190, China}
\author{G. Juzeli\={u}nas}
\email{gediminas.juzeliunas@tfai.vu.lt}\affiliation{Institute of
Theoretical Physics and Astronomy, Vilnius University, A.
Go\v{s}tauto 12, Vilnius 01108, Lithuania}
\author{An-Chun Ji}
\email{andrewjee@sina.com}\affiliation{Department of Physics,
Capital Normal University, Beijing 100048, China}
\date{{\small \today}}

\begin{abstract}
Motivated by  a goal of realizing spin-orbit coupling (SOC) beyond
one-dimension (1D), we propose  and analyze a method to generate  an
effective 2D SOC in bilayer BECs with  laser-assisted inter-layer
tunneling. We show that  an interplay between the inter-layer
tunneling,  SOC and intra-layer atomic interaction can give rise to
diverse  ground state configurations.  In particular, the system
undergoes  a transition to a new type of stripe phase which
spontaneously breaks the time-reversal symmetry. Different from the
ordinary Rashba-type SOC,  a fractionalized skyrmion lattice emerges
spontaneously in the bilayer system without  external traps.
Furthermore, we predict the occurrence of a tetracritical point in
the phase diagram of the bilayer BECs, where four different phases
merge together. The origin of the emerging different phases is
elucidated.
\end{abstract}

%67. Quantum fluids and solids
 %67.85 Ultracold gases
   % 67.85.Lm  Degenerate Fermi gases
   % 67.85.-d   Ultracold gases, trapped gases

%03. Quantum mechanics, field theories
  %03.75 matter waves of ultracold  gases
     % 03.75.Ss Degenerate Fermi gases
     % 03.75.Mn Multicomponent condensates; spinor condensates

% 30. ATOMIC AND MOLECULAR PHYSICS
  % 37.10. Atom, molecule, and ion cooling methods
     % 37.10.Jk Atoms in optical lattices

%70. CONDENSED MATTER: ELECTRONIC STRUCTURE, ELECTRICAL, MAGNETIC
    %71. Electronic structure of bulk materials
       % 71.10.Fd Lattice fermion models (Hubbard model, etc.)

%05. Statistical physics
  % 05.30.Fk Fermion systems and electron gas
  % 05.30.Jp    Boson systems

%71.70.Ej Spin Orbit coupling, Zeeman and Stark splitting

\maketitle

\section{Introduction}

The search for new exotic matter states \cite{Hasan,Qi} and the
study of phase transitions \cite{Sachdev} are currently amongst the
main issues  in the condensed matter community. During the last few
years these topics have gained an increasing interest for ultracold
atomic gases \cite{Lewenstein,Bloch,Dalibard,Lewenstein1,Goldman}
which represent the systems simulating many condensed matter
phenomena. Recent experimental progress  in the spin-orbit coupling
(SOC) of degenerate atomic gases \cite{Lin,ZhangJY,WangPJ,Cheuk,Qu}
has stimulated the theoretical studies of diverse  new  phases due
to the SOC \cite{Goldman,Zhai,Galitski,Zhou1,Zhai1},  such as
emergence of the stripe phase in atomic Bose-Einstein condensates
(BECs) \cite{WangCJ,Ho,XuZF1,Zhang,Li},  or formation of
unconventional bound states \cite{Vyasanakere,Hu,Yu1,Gong} and
topological superfluidity \cite{Gong1} for atomic fermions. It was
demonstrated that for the spin-orbit (SO) coupled  BECs, the
half-vortex (meron) ground states may develop in  harmonic traps
\cite{Sinha,Hu1,Wilson,Kawakami,Chen-unpubl}. Such topological
objects are of special interest for studying the nontrivial spin
configurations in condensed matter physics \cite
{Mhlbauer,Yu,Schmeller,Volovik}.  The Rashba-like SOC has also been
predicted for exciton-polaritons or cavity photons \cite{Tercas}.

So far, only  a special case of an equal weight of Rashba and
Dresselhaus SOC representing the 1D SOC of the form $\propto
k_x\sigma_x$ (or $\propto k_y\sigma_x$) can be realized
experimentally \cite{Lin,ZhangJY,WangPJ,Cheuk,Qu}, making the above
rich physics unaccessible in experiments. There have been many
proposals for generating 2D (3D) SOC for ultracold atoms
\cite{Ruseckas,Stanescu,Jacob,Juzeliunas1,Juzeliunas,
Campbell,XuZF2,Anderson1,Anderson2,XuZF3,Goldman,LiuLawNg}, but its
experimental realization remains a challenge.
\begin{figure}
\includegraphics[width= 0.45\textwidth]{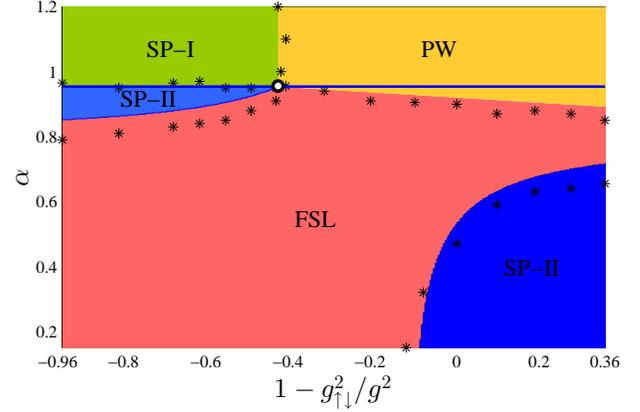}
\caption{(color online) Phase diagram of the system as  a function
of the dimensionless inter-layer tunneling $\alpha\equiv J/E_\kappa$
 ($E_\kappa=\kappa^2/2$ with $\kappa$ being the strength of SOC)
and  $1-g_{\uparrow\downarrow}^2/g^2$ ($g_{\uparrow\downarrow}/g$ is
relative atomic interaction). Here, the dimensionless intra-layer
coupling is set to be $\beta\equiv\Omega/E_\kappa=0.3$. The stars
represent the phase boundaries determined from the numerical
simulations.  A tetracritical point (TP) marked by  a circle occurs
on the critical line $\alpha^2+\beta^2=1$ (the horizontal solid
line). The colored regions are determined by variational
calculations, denoting normal stripe (SP-I, green) and plane-wave
(PW, yellow) phases,  a new type of stripe phase  (SP-II, blue), as
well as  a fractionalized skyrmion lattice (FSL, red) phase.
\label{phase1}}
\end{figure}

In this paper we propose  a  realistic way to generate an effective
2D SOC in bilayer BECs by combining current  experimental techniques
of intra-layer Raman transition \cite{Lin,ZhangJY,WangPJ,Cheuk,Qu}
and inter-layer laser-assisted tunneling
\cite{Aidelsburger,Aidelsburger1,Miyake,Atala}. The atoms in each
layer are affected by the 1D SOC in a different direction, along the
$\hat{x}$ and $\hat{y}$ axis respectively. The chiral states of
individual layers are then mixed by the laser-assisted inter-layer
tunneling, effectively providing a 2D SOC with four minimum chiral
states. Although the bilayer system bears the key properties of 2D
SOC, it is not the ordinary 2D SOC of the Rashba or Dresselhaus
type.  This give rise to a diverse phase diagram with intriguing new
matter states not encountered before.
\begin{figure}[t]
\includegraphics[width=0.48\textwidth]{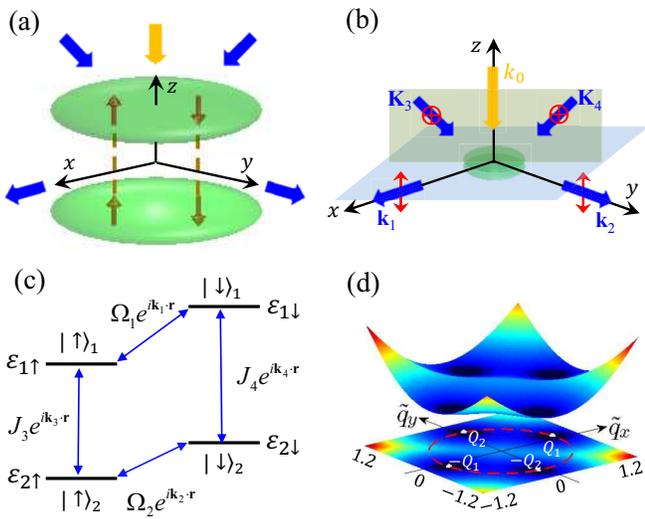}
\caption{(color online) (a) Schematic diagram of the  bilayer system
affected by a circular polarized laser field (marked in yellow)
propagating along the quantization axis $\hat{z}$, as well as four
linear polarized laser beams (marked in blue). The beams illuminate
both layers containing the atoms characterized by two internal
states $\gamma=\uparrow,\downarrow$.  (b) Illustration of a specific
laser configuration. The first and second blue laser beams are
polarized linearly along the $\hat{z}$, and propagate along the
$\hat{x}$ or $\hat{y}$ Cartesian axes. The polarizations and
frequencies of the yellow and blue beams are chosen such that they
selectively induce the Raman transition between the atomic internal
states in one of the layers. The third and fourth laser blue beam
are linearly polarized along the $\hat{x}+\hat{y}$ direction,
causing a selective laser-assisted tunnelling between the layers for
atoms in a specific atomic internal state. (c) Schematic diagram of
the intra-layer spin-flip transitions and inter-layer transitions
for specific spin states. For each layer the Raman transitions are
characterized by the coupling strength $\Omega_{1,2}$ and the recoil
momenta $\mathbf{k}_{1,2}$ along  the $\hat{x}$ and $\hat{y}$ axis
respectively. The laser induced inter-layer tunneling is
characterized by the effective strength $J_{3,4}$ and the
corresponding recoil momentum  $\mathbf{k}_{3,4}$ in the $xy$ plane.
(d) Spectrum of the single particle Hamiltonian $\hat{H}_{\rm eff}$,
Eq. (\ref{Heff}), for the relative inter-layer tunneling
$\alpha\equiv J/E_\kappa=0.6$ and intra-layer coupling
$\beta\equiv\Omega/E_\kappa=0.3$,  measured in the units of the
recoil energy $E_\kappa=\kappa^2/2$ corresponding to the momentum
$\kappa=| \mathbf{k}_{1,2} | /2$.  In that case the lowest
dispersion band has four degenerate minima at $\pm\mathbf{Q}_{1}$
and $\pm\mathbf{Q}_{2}$,  as demonstrated in the Appendix A.}
\label{Schematic}
\end{figure}

Our main  findings are summarized in Fig. \ref{phase1}.  For a large
inter-layer tunneling, the two layers are strongly coupled,  so the
usual stripe (SP-I)  or plane-wave (PW) phases  appear. For a
moderate tunneling, the system develops a new type of  the stripe
phase (SP-II), which chooses spontaneously a pair of asymmetric wave
vectors and breaks the time-reversal (TR) symmetry.  Finally, a
fractionalized skyrmion lattice (FSL) emerges  spontaneously  in the
{\it ground state}  of a {\it homogeneous} system  for a wide range
of parameters.   Such a spontaneous generation of skyrmions differs
from other  ways of their production, including thermal quenching
\cite{Su},  phase-imprinting \cite{Ruostekoski,Choi},  as well as
using  trapped systems \cite{Sinha,Wilson,Kawakami,Chen-unpubl}.

Significantly, we demonstrate that a tetracritical point (TP) occurs
among the four different phases.  The TP is a fundamental aspect in
phase transitions and has attracted  a wide interest \cite{Chaikin}.
It was first found in anisotropic antiferromagnets
\cite{Fisher,Rohrer} but has never been predicted for ultracold
atoms.

The  paper is organized as follows.  In the following Section we
introduce the general formulation for the tunneling-assisted SOC
 and discuss the single-particle spectrum.  Subsequently, in
Sec. III, we present the calculational methods by including atomic
interactions and analyze the  many-body ground state configurations
in the phase diagram. Finally in Sec. IV, we discuss some
experimental related issues and present conclusions. Details of some
derivations are presented in two Appendices.

\section{General formulation of single-body problem}
\subsection{Bilayer system}
The system under investigation is depicted in  Fig.
\ref{Schematic}(a), where an atomic BEC is confined in a bilayer
geometry.  The atoms are characterized by two internal  (quasi-spin)
states labeled by the index $\gamma=\uparrow,\downarrow$. These can
be, e.g., two magnetic sub-levels of the $F=1$ ground state manifold
of the $^{87}$Rb-type alkali atoms  \cite{Lin}  or a spin-singlet
ground state and a long-lived spin triplet excited state of the
alkaline-earth atoms \cite{Gerbier}.  In the following discussion,
we shall concentrate on the former case. However,  the results
obtained can be applied also to other systems.

The atoms are trapped by a double-well like optical potential along
the $z$ direction, but their motion is not confined in the $xy$
plane. The single-particle Hamiltonian is given by
\begin{equation}
\hat{H}_{0}=\hat{H}_{\rm atom}+\hat{H}_{\rm LIT} +\hat{H}_{\rm
LAT}\,,\label{eq:h0}
\end{equation}
where the first term  $\hat{H}_{\rm atom}$ corresponds to the
unperturbed atomic motion, the other two terms being due to the
laser induced intra-layer transitions (LIT) between the two atomic
internal states, as well as the  laser assisted tunneling (LAT) of
atoms between two wells without changing an atomic internal state.

\subsection{Atomic Hamiltonian}
The atomic Hamiltonian reads in the second quantized representation
\begin{eqnarray}
\hat{H}_{\rm atom}\!\!\!&=&\!\!\!\!\int\!\!
d^{2}\mathbf{r}dz\!\!\!\sum_{\gamma=\uparrow,\downarrow}\!\!
\hat{\psi}_{\gamma}^{\dag}(\mathbf{r},z)
\!\!\left[-\frac{\nabla_{\mathbf{r}}^{2}
+\nabla_{z}^{2}}{2}+\!\!V_{{\rm
op}}(z)\right]\!\!\hat{\psi}_{\gamma}(\mathbf{r},z)\,,\nonumber\\
\label{h01}
\end{eqnarray}
where $\hat{\psi}_{\gamma}(\mathbf{r},z)$ is an operator for
annihilation of an atom positioned at
$\mathbf{R}=\mathbf{r}+\hat{z}z$
%$\mathbf{R}\equiv\{\mathbf{r},z\}$
and in an internal state $\gamma$. Here $\mathbf{r}$ is the 2D
radius vector describing the atomic motion within a layer in the
$xy$ plane, and the coordinate $z$ characterizes the inter-layer
motion. Here also $V_{{\rm op}}$ is a  double-well optical potential
along the $z$ axis. For instance, it can be taken to be a sum of two
inverted Gaussians: $V_{{\rm op}}(z)=-V_0e^{-(z-d/2)^2}-\eta V_0
e^{-(z+d/2)^2}$ \cite{Fortanier}, where $V_0$ is  a depth of the
potential and $\eta$ is  an asymmetry parameter.

Assuming that the atoms are tightly confined in individual wells
forming the asymmetric double-well, one can expand the field
operator entering Eq. (\ref{h01}) as  \cite {Smerzi,Raghavan}:
\begin{equation}
\hat{\psi}_{\gamma}(\mathbf{r},z) =\hat{\psi}_{1\gamma}(\mathbf{r})
\phi_{1}(z)+\hat{\psi}_{2\gamma}(\mathbf{r})
\phi_{2}(z)\,,\label{expansion}
\end{equation}
where $\hat{\psi}_{j\gamma}(\mathbf{r})$ represents an operator for
annihilation of an atom in the $j$-th layer  and internal state
$\gamma$. The functions $\phi_{1,2}(z)$ describe two states
localized at an individual layer for the atomic motion along the $z$
axis. They can be constructed by  taking a superposition of the
symmetric $\Phi_+$ and antisymmetric $\Phi_-$  atomic eigenstates,
$\phi_{1,2}(z)=(\Phi_+\pm\Phi_-)/\sqrt{2}$,  for a completely
symmetric double-well system
 \cite {Raghavan} corresponding
to $\eta =1$ in  $V_{{\rm op}}(z)$.  Such states are normalized and
orthogonal to each other (analogous to the Wannier states in a
periodic potential), and are characterized by the lowest energies
$\varepsilon_{j=1,2}$ of each well.

Using Eq. (\ref{expansion}) for $\hat{\psi}_{\gamma}(\mathbf{r},z)$
and integrating over $z$ in Eq. (\ref{h01}), the two-layer
Hamiltonian takes the form
\begin{eqnarray}
\hat{H}_{\rm atom}=\int
d^{2}\mathbf{r}\sum_{j=1,2;\gamma=\uparrow,\downarrow}
\hat{\psi}_{j\gamma}^{\dag}(\mathbf{r})\left[\frac{\mathbf{q}^{2}}{2}
+\varepsilon_{j}\right]\hat{\psi}_{j\gamma}(\mathbf{r})\,.\label{h02}
\end{eqnarray}
where $\mathbf{q}=-i\hbar\nabla_{\mathbf{r}}$ is the momentum
operator for the atomic motion in the $xy$ plane, and the lowest
energy of atoms in each well is given by \cite {Raghavan}
\begin{equation}
\varepsilon_{j}=\int
dz\phi_{j}^*(z)\left[-\frac{\nabla_{z}^{2}}{2}+V_{{\rm
op}}(z)\right] \phi_{j}(z)\,. \label{eigen-equation}
\end{equation}
Note that generally there should be a tunneling matrix element
$\mathcal{K}=\int
dz\phi_{1}^*(z)\left[-\frac{\nabla_{z}^{2}}{2}+V_{{\rm
op}}(z)\right] \phi_{2}(z)$ between two layers  in Eq. (\ref{h02}).
However,  for a  sufficiently asymmetric double-well potential
\cite{Atala,Porto},  the inter-layer coupling is small compared with
energy mismatch between the wells.  As a result, the direct
inter-layer tunneling is inhibited and hence can be neglected.  In
this case, the wave-functions $\phi_{j=1,2}(z)$  localized  on
individual wells become nearly  true eigenstates of the full
asymmetric double-well potential.

In the following, the double-well potential is assumed to be
state-dependent: $V_{{\rm op}}(z)\equiv V_{{\rm op}}^{(\gamma)}(z)$.
Thus  one should replace the lowest energy $\varepsilon_{j}$ by a
state-dependent energy $\varepsilon_{j\gamma}$  in Eq. (\ref{h02}).
The state-dependence of the double-well potential can be implemented
e.g. by making use of a Zeeman shift that varies along the $\hat{z}$
axis due to the magnetic field gradient \cite{Lin2} or by
additionally applying cross-polarized laser fields
counter-propagating along the $\hat{z}$ axis to yield a
state-dependent optical lattice along that direction
\cite{Deutsch,McKay}.

\subsection{Atom-light interaction}

Now, we turn to the atom-light interaction processes which induce
both the intra-layer SOC and also inter-layer tunnelling. For this
let us present a general Hamiltonian $H_{\rm AL}^{\left(\rm
full\right)}$ for the atom-light coupling in an atomic hyperfine
ground-state manifold
%with arbitrary fixed value of the total atomic spin
described by the total spin operator $\hat{\mathbf{F}}$.
It can be represented in
terms of the scalar and vector light shifts \cite{Goldman,Deutsch}:
\begin{align}
H_{\rm AL}^{\left(\rm full\right)}= & \
u_{s}(\mathbf{E}^{*}\!\cdot\!\mathbf{E}) +\frac{iu_{{v}}g_{F}}{\hbar
g_{J}}\left(\mathbf{E}^{*}\times\mathbf{E}\right)
\cdot\hat{\mathbf{F}}\,,\label{eq:light-potential}
\end{align}
where $\mathbf{E}$ is a negative frequency part of the full
 electric field,
%$\hat{\mathbf{F}}$ is the full atomic spin operator,
$u_{s}$ and $u_{v}$ are the scalar and vector atomic
polarizabilities with $u_{s}\gg u_{v}$  for the detuning exceeding
the fine-structure splitting of the excited electronic state. Here
also $g_{J}$ and $g_{F}$ are the Land\'{e} g-factors for the
electronic spin and the total angular momentum of the atom,
respectively. In the case of the $^{87}\rm Rb$ atom we have
$g_{F}/g_{J}=-1/4$ for the lowest energy hyperfine manifold with
$F=1$.

Figures \ref{Schematic}(a,b) illustrate a possible laser
configuration implementing the required intra- and inter-layer
coupling.  As shown in Fig.~\ref{Schematic}(a), both layers are
simultaneously illuminated by a circular polarized laser field
(marked in yellow) propagating along the quantization axis $\hat{z}$
with the electric field $\mathbf{E}_0\sim
(\hat{x}+i\hat{y})e^{i(k_0z-\omega_0 t)}$,  as well as by four
linear polarized laser beams, $\mathbf{E}_{j}$ with $j=1,2,3,4$
(marked in blue).

\subsubsection{Intra-layer transitions}
The first and second  blue laser beams ($j=1,2$)  take care of the
 intra-layer transitions.  They are characterized by the electric field
 $\mathbf{E}_{j}\sim \hat{z}e^{i[\mathbf{k}_{j}\cdot\mathbf{r}-(\omega_0+\delta\omega_{j})
t]}$ polarized linearly along $\hat{z}$,  and contain wave-vectors
$\mathbf{k}_{1}$ and $\mathbf{k}_{2}$  oriented along the $\hat{x}$
and $\hat{y}$ Cartesian axes, respectively, see Fig.
\ref{Schematic}(b).

The  frequencies of the fields $\mathbf{E}_0$ and  $\mathbf{E}_{j}$
satisfy the two-photon resonance condition for the intra-layer
transitions between the atomic internal states. Specifically we
have: $\delta\omega_{j}=\Delta_j$ with $j=1,2$, where
$\Delta_j\equiv\varepsilon_{j\downarrow} -\varepsilon_{j\uparrow}$
is the energy of the Zeeman splitting between atomic internal states
$|m_F = -1\rangle\equiv|\downarrow\rangle$ and $|m_F =
0\rangle\equiv|\uparrow\rangle$  in the $j$-th layer. Due to a
sufficiently large quadratic Zeeman effect field, the $|m_F =
1\rangle$ magnetic sublevel is out of the Raman resonance and hence
can be ignored, like in the initial NIST experiment on the SOC for
ultracold $^{87}\rm Rb$ gases  \cite{Lin}. Thus  the field
$\mathbf{E}_0$  together with  $\mathbf{E}_{j}$ selectively induce
the Raman transition between the atomic internal states $|m_F =
-1\rangle\equiv|\downarrow\rangle$ and $|m_F =
0\rangle\equiv|\uparrow\rangle$  in $j$-th layer, as schematically
depicted in Fig. \ref{Schematic}(c). They are represented by the
second term in Eq. (\ref{eq:light-potential}) with
$\mathbf{E}^*_{1,2}\times \mathbf{E}_0\sim (\hat{x}+i\hat{y})$,
which gives rise to atomic spin-flip transitions. In this way, the
Hamiltonian describing the laser induced intra-layer transitions
reads
\begin{equation}
\hat{H}_{\rm LIT} =\int d^{2}\mathbf{r}\sum_{j=1,2}
\left[\Omega_{j}e^{i\varphi_j}+c.c.\right]
\hat{\psi}_{j\uparrow}^{\dag}(\mathbf{r})\hat{\psi}_{j\downarrow}(\mathbf{r})
+{\rm H.c.}\,, \label{H2}
\end{equation}
with $\varphi_j = \mathbf{k}_{j}\cdot\mathbf{r}-\delta\omega_{j}t$,
where $\Omega_{j}$ denotes the   Rabi frequency of the Raman
coupling. Since the the bilayer potential strongly confines the
atomic motion in the $xy$ plane, the out of plane recoil momentum
$-k_{0}\hat{z}$ is not important for the intra-layer transitions and
hence does not show up in the Hamiltonian  (\ref{H2}).

\subsubsection{Inter-layer tunneling}
The third and fourth (blue) laser  beams are linearly polarized
along the $\hat{x}+\hat{y}$  direction with $\mathbf{E}_{3,4}\sim
(\hat{x}+\hat{y})e^{i[\mathbf{K}_{3,4} \cdot
\mathbf{R}-(\omega_0+\delta\omega_{3,4}) t]}$, where
$\mathbf{K}_{3,4}=\mathbf{k}_{3,4}+\hat{z}k_z$
%and$\mathbf{R}=\mathbf{r}+\hat{z}z$
are the 3D wave-vectors. Their in-plane components
$\mathbf{k}_{3}=(\mathbf{k}_{2}-\mathbf{k}_{1})/2$ and
$\mathbf{k}_{4}=(\mathbf{k}_{1}-\mathbf{k}_{2})/2$  match with the
corresponding wave-vectors $\mathbf{k}_{1}$ and $\mathbf{k}_{2}$ for
the intra-layer transitions. This yields a zero in-plane momentum
transfer for the atomic transitions over the closed loop shown in
Fig. \ref{Schematic}(c):
$\mathbf{k}_{1}-\mathbf{k}_{2}+\mathbf{k}_{3}-\mathbf{k}_{4}=0$.
Since the frequencies of all the laser beams inducing Raman
transitions are very close to each other, we have
$|\mathbf{k}_{1}|\approx|\mathbf{k}_{2}|\approx|\mathbf{K}_{3}|
\approx|\mathbf{K}_{4}|\approx |k_{0}|$. Consequently the matching
condition for the in-plane wave-vectors implies that $k_z\approx \pm
k_{0}/\sqrt{2}$.

The fields $\mathbf{E}_0$  and  $\mathbf{E}_{3,4}$ are not
orthogonal to each other $\mathbf{E}^*_{3,4} \cdot \mathbf{E}_0\ne
0$, and hence provide a scalar light shift represented by the first
term in Eq. (\ref{eq:light-potential}). It oscillates with the
frequency $\delta\omega_{3,4}$ and enables the inter-layer
transitions \cite{Aidelsburger,Aidelsburger1,Miyake,Atala}. The
frequencies of the laser beams are assumed to satisfy the conditions
of the two-photon inter-layer resonance,
$\delta\omega_{3}=\Delta_{\uparrow}$ and
$\delta\omega_{4}=\Delta_\downarrow$ for each internal state $\gamma
= \uparrow\,,\downarrow$, where
$\Delta_\gamma\equiv\varepsilon_{1\gamma}-\varepsilon_{2\gamma}$.
This ensures a {\it selective laser-assisted tunnelling} between the
layers for atoms in a specific atomic internal state, as
schematically depicted in Fig. \ref{Schematic}(c).

In this way, the Hamiltonian describing the laser-assisted
tunnelling is given by
\begin{eqnarray}
\hat{H}_{\rm LAT} &=&\int d^{2}\mathbf{r} \left(J_{3}e^{i\varphi_3}
+c.c.\right) \hat{\psi}_{2\uparrow}^{\dag}(\mathbf{r})
\hat{\psi}_{1\uparrow}(\mathbf{r})
+{\rm H.c.}\nonumber\\
&+&\int d^{2}\mathbf{r} \left(J_{4}e^{i\varphi_4} +c.c.\right)
\hat{\psi}_{2\downarrow}^{\dag}(\mathbf{r})
\hat{\psi}_{1\downarrow}(\mathbf{r})
+{\rm H.c.}\,,\nonumber\\
\label{H3}
\end{eqnarray}
where
$\varphi_{3,4}=\mathbf{k}_{3,4}\cdot\mathbf{r}-\delta\omega_{3,4}t$.
The strength of the inter-layer coupling $J_{3,4}=\Omega_{3,4}\int
dz\phi_{2}^{*}(z)\phi_{1}(z)e^{ik_{z}^\prime z}$ depends on the Rabi
frequency $\Omega_{3,4}$ of the atom-light coupling, the overlap of
the wave-functions $\phi_{1}(z)$ and $\phi_{2}(z)$ of individual
wells, as well as the $z$ component of the momentum transfer
$k_{z}^\prime=k_{z}-k_0$. The latter equals to $k_{z}^\prime=\pm
k_0\sqrt{2}/2-k_0$ depending on the sign of $k_z=\pm k_0/\sqrt{2}$.

As discussed in the paragraph following Eq. (\ref{expansion}), the
states localized on each layer $\phi_{1}(z)$ and $\phi_{2}(z)$ are
the Wannier state analogs for the double well potential.  These
states are orthogonal, so it is the factor $e^{ik_z^\prime z}$ due
to the momentum transfer along the tunnelling direction $\hat{z}$
which makes the overlap integral $J_{3,4}$ non-zero \cite{Miyake}.

\subsection{Elimination of the spatial and temporal
dependence of the Hamiltonian $\hat{H}_{0}$}

To eliminate the spatial and temporal dependence in the
single-particle Hamiltonian $\hat{H}_{0}$,  we perform a unitary
transformation $\hat{U}=e^{-i\hat{S}}$, with
\begin{eqnarray}
\hat{S} =\int
d^{2}\mathbf{r}\!\!\!\!\sum_{j=1,2;\gamma=\uparrow,\downarrow} \!\!
\left(\varepsilon_{j\gamma}t+m_{\gamma}\mathbf{k}_{j}
\cdot\mathbf{r}\right)\hat{\psi}_{j\gamma}^{\dag}(\mathbf{r})
\hat{\psi}_{j\gamma}(\mathbf{r})\,,\label{U}
\end{eqnarray}
where $m_{\uparrow}=1/2$ and  $m_{\downarrow}=-1/2$. The Hamiltonian
$\hat{H}_{0}$ transforms to
$\hat{H}_{0}^\prime=\hat{U}^{\dagger}\hat{H}_{0}\hat{U}
-i\hbar\hat{U}^{\dagger}\partial_{t}\hat{U}$, where the last term,
due to the time-dependence of $\hat{U}$, eliminates the energies
$\varepsilon_{j\gamma}$ featured in the Hamiltonian $\hat{H}_{\rm
atom}$, Eq. (\ref{h02}). The transformed operators entering
$\hat{H}_{\rm LIT}$ and $\hat{H}_{\rm LAT}$ acquire extra time- and
position-dependent factors:
$\hat{U}^{\dagger}\hat{\psi}_{j\uparrow}^{\dag}(\mathbf{r})
\hat{\psi}_{j\downarrow}(\mathbf{r})\hat{U}
=\hat{\psi}_{j\uparrow}^{\dag}(\mathbf{r})
\hat{\psi}_{j\downarrow}(\mathbf{r})e^{i[\left(\varepsilon_{j\downarrow}
-\varepsilon_{j\uparrow}\right)t-\mathbf{k}_{j}\cdot\mathbf{r}]}$
and $\hat{U}^{\dagger}\hat{\psi}_{2\gamma}^{\dag}(\mathbf{r})
\hat{\psi}_{1\gamma}(\mathbf{r})\hat{U}=\hat{\psi}_{2\gamma}^{\dag}(\mathbf{r})
\hat{\psi}_{1\gamma}(\mathbf{r})e^{i[\left(\varepsilon_{1\gamma}
-\varepsilon_{2\gamma}\right)t+m_{\gamma}\left(\mathbf{k}_{1}-\mathbf{k}_{2}\right)
\cdot\mathbf{r}]}$.

In what follows we use the resonance conditions for the intra- and
inter-layer laser-induced transitions, and apply the rotating wave
approximation to ignore the fast oscillating terms in the
transformed Hamiltonian.  This is legitimate if the inter- and
intra-layer detunings exceed the corresponding transition matrix
elements. As a result, one arrives at the following time- and
position-independent single-particle Hamiltonian:
\begin{equation}
\hat{H}_{0}^\prime =\hat{H}_{\rm atom}^\prime+\hat{H}_{\rm
LIT}^\prime +\hat{H}_{\rm LAT}^\prime\,,
\label{H_0-prime}
\end{equation}
with
\begin{eqnarray}
\!\!\!\!\hat{H}_{\rm atom}^\prime\!\!&=&\!\!\int
d^{2}\mathbf{r}\!\!\!\!\sum_{j=1,2;\gamma=\uparrow,\downarrow}\!\!\!\!
\hat{\psi}_{j\gamma}^{\dag}(\mathbf{r})
\frac{\left(\mathbf{q}+m_{\gamma}\mathbf{k}_{j}\right)^{2}}{2}
\hat{\psi}_{j\gamma}(\mathbf{r})\,,\label{H00}\\
\!\!\!\! \hat{H}_{\rm LIT}^\prime\!\!& = & \!\!\int
d^{2}\mathbf{r}\sum_{j=1,2}\Omega_{j}
\hat{\psi}_{j\uparrow}^{\dag}(\mathbf{r})
\hat{\psi}_{j\downarrow}(\mathbf{r}) +{\rm H.c.}\,,\label{H2-1}
\end{eqnarray}
and
\begin{eqnarray}
\hat{H}_{\rm LAT}^\prime=\int
d^{2}\mathbf{r}\sum_{\gamma=\uparrow,\downarrow} J_{\gamma}
\hat{\psi}_{2\gamma}^{\dag}(\mathbf{r})
\hat{\psi}_{1\gamma}(\mathbf{r})+{\rm H.c.}\,.\label{H3-1}
\end{eqnarray}
Here we have made use of the matching condition for in-plane
wave-vectors:
$\mathbf{k}_{\gamma}=m_\gamma\left(\mathbf{k}_{2}-\mathbf{k}_{1}\right)$,
with $\gamma=\uparrow,\downarrow$, $\mathbf{k}_{\uparrow}\equiv
\mathbf{k}_{3}$ and
 $\mathbf{k}_{\downarrow}\equiv \mathbf{k}_{4}$.
This enables us to remove the position-dependent phase factors
$e^{i\left[\mathbf{k}_{\gamma}\cdot\mathbf{r}
+m_{\gamma}\left(\mathbf{k}_{1}
\!-\!\mathbf{k}_{2}\right)\cdot\mathbf{r}\right]}$ in
Eq.~(\ref{H3-1}) for
 $\hat{H}_{\rm LIT}^\prime$.
In Eq.(\ref{H3-1}) we have also rewritten $J_{3,4}$  as
$J_{\uparrow,\downarrow}$. In general, $\Omega_{j}$ and $J_{\gamma}$ are independent complex variables with tunable relative phases. In what follows we take them to be real. In that case one needs to stabilise properly the phases of the laser beams inducing the 
atomic inter-layer tunnelling and intra-layer transitions. The phase stabilisation is experimentally challenging but feasible;
it has been done in a recent experiment on the two-component slow light \cite{Lee2014NCom}. 

It is convenient to introduce two-component  row and column bosonic
field operators  for creation and annihilation of an atom in the
$j$-th layer:
$\hat{\psi}_{j}^{\dag}(\mathbf{r})=[\hat{\psi}_{j\uparrow}^{\dag}
(\mathbf{r}),\hat{\psi}_{j\downarrow}^{\dag}(\mathbf{r})]$ and
$\hat{\psi}_{j}(\mathbf{r})=[\hat{\psi}_{j\uparrow}(\mathbf{r}),
\hat{\psi}_{j\downarrow}(\mathbf{r})]^{T}$.  Omitting a constant
term, the full single particle Hamiltonian (\ref{H_0-prime}) can
then be represented as
\begin{eqnarray}
\hat{H}_{0}^\prime & = & \int d^{2}\mathbf{r}\sum_{j=1,2}
\hat{\psi}_{j}^{\dag}(\mathbf{r})\left[\frac{\mathbf{q}^{2}+\mathbf{q}
\cdot\mathbf{k}_{j}\sigma_{z}}{2m}+\Omega_{j}\sigma_{x}\right]
\hat{\psi}_{j}(\mathbf{r})\nonumber \\
& + & \int d^{2}\mathbf{r}\sum_{\gamma=\uparrow,\downarrow}
\left(J_{\gamma}\hat{\psi}_{1\gamma}^{\dag}(\mathbf{r})
\hat{\psi}_{2\gamma}(\mathbf{r})+{\rm H.c.}\right)\,.\label{AHeff}
\end{eqnarray}

We assume that the coupling strengths are state- and
site-independent ($J_{\gamma}=J$ and $\Omega_{j}=\Omega$). Since the
wave-vectors $\mathbf{k}_{1}$ and $\mathbf{k}_{2}$ are oriented
along the $\hat{x}$ and $\hat{y}$ Cartesian axes, the SOC in each
layer is along these directions:
$\mathbf{q}\cdot\mathbf{k}_{j}=q_{j}k_{j}=2q_{j}\kappa_{j}$, with
$q_{1}=q_{x}$ and $q_{2}=q_{y}$. Here  $\kappa_j=|\mathbf{k}_j|/2
\approx |k_0|/2 \equiv  \kappa $ denotes the strength of  the SOC
which is the same for both layers $j=1,2$. Interchanging the spin
operators, $\sigma_{x}\rightarrow-\sigma_{z}$ and
$\sigma_{z}\rightarrow\sigma_{x}$, one arrives at the effective
single-particle second-quantized Hamiltonian
\begin{eqnarray}
\hat{H}_{\rm eff}&=&\int d^2
\mathbf{r}\sum_{j=1,2}\hat{\psi}_j^\dag(\mathbf{r})
\left[\frac{\mathbf{q}^2}{2} +\kappa
q_j\sigma_x-\Omega\sigma_z\right]\hat{\psi}_j
(\mathbf{r})\nonumber\\
&+&J\int d^2 \mathbf{r}\sum_{\gamma=\uparrow,\downarrow}
\left(\hat{\psi}_{1\gamma}^\dag(\mathbf{r})
\hat{\psi}_{2\gamma}(\mathbf{r})+{\rm H.c.}\right)\,.\label{Heff}
\end{eqnarray}

Finally we define the dimensionless  momentum
$\widetilde{\mathbf{q}}\equiv\mathbf{q}/\kappa$,  the dimensionless
energies of the intra-layer coupling $\beta\equiv\Omega/E_\kappa$ as
well as the inter-layer tunneling $\alpha\equiv J/E_\kappa$ measured
in the units of the  energy $E_\kappa=\kappa^2/2$.

\subsection{Single-Particle Spectrum}

\begin{figure}
\setlength{\abovecaptionskip}{-0.5cm}
\centerline{\includegraphics[width= 0.5\textwidth]{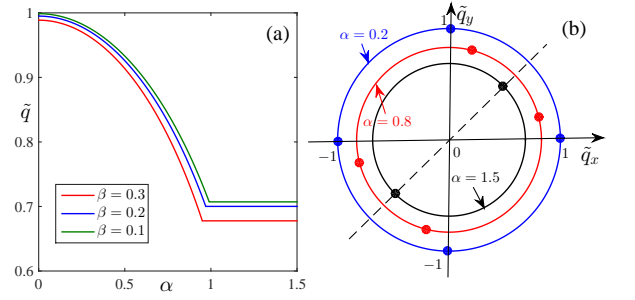}}
\caption{(color online) (a) The momentum evolution of the single particle ground state with the interlayer tunneling $\alpha$ for fixed intra-layer coupling $\beta$. (b) 
Energy minima (denoted by solid circles) of the single particle ground state in the plane of the dimensionless momentum ($\tilde{q}_x$-$\tilde{q}_y$) for  $\beta=0.1$  and different values of the interlayer tunneling $\alpha$. Note that for each $\alpha$ all the energy minima are situated on the same circle with a radius $\tilde{q}$.
\label{momentum}}
\end{figure}

For bosonic systems, the single-particle states play an important
role in determining the ground state configurations. In Appendix A,
we introduce a single spinor
$\hat{\Psi}(\mathbf{r})=[\hat{\psi}_{1\uparrow}
(\mathbf{r}),\hat{\psi}_{1\downarrow}(\mathbf{r}),\hat{\psi}_{2\uparrow}
(\mathbf{r}),\hat{\psi}_{2\downarrow}(\mathbf{r})]^T$ to treat the
double layer.
%and rewrite the Hamiltonian (\ref{Heff}) in
%combination with the notation $\sigma$ and $\tau$ representing the
%spin and layer index respectively.
In the four-component basis, the single-particle spectrum possesses
four branches, and here we are interested in the lowest branch of
energy spectrum, as depicted in Fig. \ref{Schematic}(d). First for
$\alpha=0$, the two layers are decoupled, and there are two pairs of
degenerate energy minima along the $\widetilde{q}_x$ and
$\widetilde{q}_y$ directions, respectively. Then, by increasing
$\alpha$, the inter-layer tunneling  couples  the  two pairs of
minima together, resulting in the four minimum chiral states  at
$\pm\mathbf{Q}_{1}=\pm(\widetilde{q}^+_0, \widetilde{q}^-_0)$ and
$\pm\mathbf{Q}_{2}=\pm(\widetilde{q}^-_0, \widetilde{q}^+_0)$, where
$\widetilde{q}^\pm_0=\frac{1}{2}
(\sqrt{Q_0^2+\alpha^2/2}\pm\sqrt{Q_0^2-\alpha^2/2})$,  with
$Q_0=|\mathbf{Q}_{1,2}|$ satisfying a nonlinear equation  given by
Eq. (\ref{nonlinear}) in Appendix A.  The four energy minima are located on the same circle with a radius $\tilde{q}$ and satisfy a
reflection symmetry upon the diagonal axis in the $\tilde{q}_x$-$\tilde{q}_y$ plane. When  the dimensionless
inter-layer coupling $\alpha$  is increased, the momentum of the single-particle ground state  decreases monotonically as shown in Fig. (\ref{momentum}a). In particular, above the critical line with $\alpha^2+\beta^2=1$, the  energy minima of the chiral states
converge to $\pm\mathbf{Q}$, with
$\mathbf{Q}=\sqrt{1-\beta^2}(1/2,1/2)$  situated on the  diagonal
axis as shown in Fig. (\ref{momentum}b). In this case, for strong  intra-layer coupling $\beta\geq1$,
the minima shrink to  a single point at $\mathbf{Q}=\mathbf{0}$.

\section{MANY-BODY GROUND STATES}
\subsection{Calculational Methods}
Since the interaction between the atoms is short ranged, it is much
stronger for the  atoms situated at the same layer than at different
layers  of the bilayer BEC.  Neglecting the inter-layer interaction,
the second-quantized Hamiltonian describing the atom-atom
interaction reads
\begin{eqnarray}
\hat{H}_{\rm{int}}=\int d^2 \mathbf{r}\sum_{j=1,2} \left(\frac{
g_\uparrow}{2}\hat{n}_{j\uparrow}^2 +\frac{
g_\downarrow}{2}\hat{n}_{j\downarrow}^2 +
g_{\uparrow\downarrow}\hat{n}_{j\uparrow}\hat{n}_{j\downarrow}\right)\,,
\label{intH}
\end{eqnarray}
where  $g_{\uparrow}$ and $g_{\downarrow}$ are the strengths of the
interaction between the atoms in the same internal (quasi-spin)
states,  $g_{\uparrow\downarrow}$ is the corresponding interaction
strength for the atoms in different internal states,
$\hat{n}_{j\gamma}=\hat{\psi}^\dag_{j\gamma}\hat{\psi}_{j\gamma}$
 being the operator for the  atomic density in the $j$-th layer
and the internal state $\gamma= \uparrow , \downarrow$. We shall
first assume the symmetric intra-species interaction with
$g_{\uparrow,\downarrow}=g$. In this paper, we consider a weakly interacting case so that the quantum fluctuations can be neglected legitimately \cite{RMP}. Under the mean-field level, the zero-temperature  ground-state structures can then be
investigated by numerically solving the  mean-field Gross-Pitaevskii
(G-P) equation  for the two-component wave-function (vector order
parameter) of the condensate $\psi_{j\gamma}\equiv
\langle\hat{\psi}_{j\gamma}\rangle$. The G-P energy functional reads
$\mathcal{E}[\bar{\psi}_{j\gamma},\psi_{j\gamma}]=\langle\hat{H}_{\rm
eff}+\hat{H}_{\rm{int}}\rangle$,  giving
\begin{eqnarray}
&&\mathcal{E}[\bar{\psi}_{j\gamma},\psi_{j\gamma}] =\int
d^2\mathbf{r}\left[\sum_{j,\gamma}{\bar\psi}_{j\gamma}\left(-\frac{1}{2}
\nabla^2+\frac{1}{2}\omega^2r^2\right)\psi_{j\gamma}\right.\nonumber\\
&&\left.+J\sum_\gamma \!\left(\bar{\psi}_{1\gamma}\psi_{2\gamma}
\!+\!\bar{\psi}_{2\gamma}\psi_{1\gamma}\right)
\!-\!i\kappa\left(\bar{\psi}_{1\uparrow}\partial_x\psi_{1\downarrow}
\!+\!\bar{\psi}_{1\downarrow}\partial_x\psi_{1\uparrow}\right.\right.\nonumber\\
&&\left.\left.+\bar{\psi}_{2\uparrow}\partial_y\psi_{2\downarrow}
+\bar{\psi}_{2\downarrow}\partial_y\psi_{2\uparrow}\right)
+\sum_j\Omega\left(|\psi_{j\uparrow}|^2-|\psi_{j\downarrow}|^2\right)\right.\nonumber\\
&&\left.+\sum_{j}\left(\frac{g_\uparrow}{2}|\psi_{j\uparrow}|^4
+\frac{g_\downarrow}{2}|\psi_{j\downarrow}|^4
+g_{\uparrow\downarrow}|\psi_{j\uparrow}|^2|\psi_{j\downarrow}|^2\right)\right]\,,\nonumber\\
\label{G-P}
\end{eqnarray}
where we have taken the BEC wave function  to be normalized to the
unity: $\int d^2\mathbf{r}
\sum_{j,\gamma}|\psi_{j\gamma}(\mathbf{r})|^2=1$.  This has been
carried out via the substitution
$\psi_{j\gamma}\rightarrow\sqrt{N}\psi_{j\gamma}$ which implies
rescaling of the interaction strengths $g_{\uparrow\downarrow}
\rightarrow N g_{\uparrow\downarrow} $, $g_{\uparrow} \rightarrow N
g_{\uparrow}$ and $g_{\downarrow} \rightarrow N g_{\downarrow}$,
where $N$ is the total number of atoms.  To deal with the BEC
confined in a finite area, in Eq. (\ref{G-P}) we have included a
sufficiently weak harmonic  trapping potential with a frequency
$\omega$  much smaller than the SOC frequency $E_\kappa$.

By minimizing Eq. (\ref{G-P}) via the imaginary time evolution
method, we  have derived various  phases as shown by the stars in
Fig. \ref{phase1}.  To reveal the underlying physics of the phases,
let us explore whether it is possible for the bilayer atoms to
condense simultaneously at two pairs of wave vectors
$\pm\mathbf{Q}_{1}$ and $\pm\mathbf{Q}_2$. First we note that the
triangular lattice phases have been found  for a trapped spin-1/2
BEC with Rashba-type SOC \cite{Sinha,Hu1}.  Furthermore, the
triangular and square lattice phases have  also been observed for a
spin-2 BEC \cite{Xu2}. Yet, for a spin-1/2 BEC in a 2D homogeneous
system, the ground states are found to be plane wave or stripe
phases comprised of a single wave vector or a pair of wave vectors,
and it is hard to form  the ground state which involves an
interference of more than one pair of wave vectors \cite{WangCJ}.
Even  if a square lattice is added to break the translational
symmetry leading to the four minimum chiral states, the  G-P ground
states still favor the normal stripe phase \cite{Cole}.   This is
because in a 2D Rashba-type system without external traps, a state
with more than one pair of wave vectors has a non-uniform density
modulation and is energetically unfavorable.
\begin{figure}
\includegraphics[width= 0.47\textwidth]{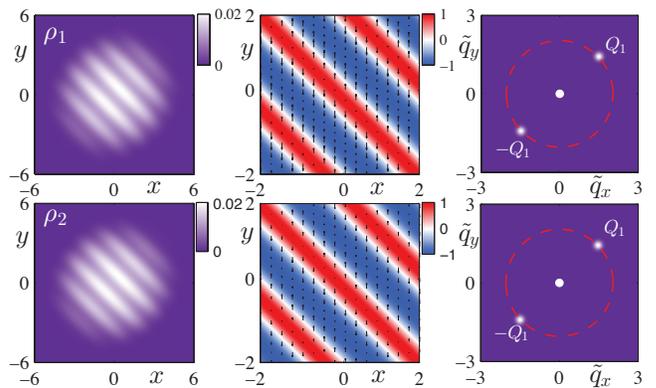}
\caption{(color online)  Total density profiles $\rho_j$ (left),
spin textures $\mathbf{S}_j$ (middle), and the corresponding
momentum distributions (right) in the first (up) and second (bottom)
layer for normal SP-I phase with $\alpha=1.1$ and
$g_{\uparrow\downarrow}/g=1.3$. The color in the spin textures
indicates the magnitude of $S_{jz}$. \label{stripe-I}}
\end{figure}

However, in the proposed bilayer system  where only  atoms situated
in the same layer attract repulsively,  it  is energetically more
favorable to delocalize the atoms in both layers. In this case, a
competition of the intra-layer atomic interactions and inter-layer
tunneling may couple the four minimum  energy states in a different
manner and lead to a number of new  phases.

To study  a possible of formation of interfering multi-wave ground
states, we take the following  Ansatz
$\psi_G\equiv\langle\hat{\Psi}\rangle=a_{1+}\psi_{+\mathbf{Q}_{1}}
+a_{1-}\psi_{-\mathbf{Q}_{1}}
+a_{2+}\psi_{+\mathbf{Q}_{2}}+a_{2-}\psi_{-\mathbf{Q}_{2}}$. Here
$\psi_{\pm\mathbf{Q}_{1,2}}\equiv\phi_{\pm\mathbf{Q}_{1,2}} e^{\pm
i\mathbf{Q}_{1,2}\cdot\mathbf{r}}$  denote  the four-component
eigen-functions corresponding to four degenerate energy minima
[given by Eq. (\ref{eigen-wave}) in the Appendix A], and $a_{1\pm}$,
$a_{2\pm}$ are complex amplitudes satisfying the normalization
condition.  The corresponding variational interacting energy
functional
$\mathcal{E}[a_{1\pm},a_{2\pm}]=\langle\hat{H}_{\rm{int}}\rangle$
reads
\begin{eqnarray}
&&\mathcal{E}[a_{1\pm},a_{2\pm}]\nonumber\\
&&=C_1\sum_{\pm}\sum_{j=1,2}|a_{j\pm}|^4+C_2
\sum_{\pm}\sum_{i\neq j}|a_{i\pm}|^2|a_{j\pm}|^2\nonumber\\
&&+C_3\sum_{\pm}\sum_{i\neq j}|a_{i\pm}|^2|a_{j\mp}|^2+C_4
\sum_{\pm}\sum_{j=1,2}|a_{j\pm}|^2|a_{j\mp}|^2\nonumber\\
&&-2|C_5||a_{1+}a_{1-}a_{2+}a_{2-}|\,,\label{variation}
\end{eqnarray}
where the coefficients $C_{1-5}$ are presented in  the Appendix B.
By minimizing the energy $\mathcal{E}[a_{1\pm},a_{2\pm}]$, we find
that all the emerging phases predicted by the numerical simulations
of G-P equations can be identified by the variational results as
shown by the colored regions in Fig. \ref{phase1}.  This provides a
deeper insight into the nature of the ground state configurations
analyzed in the following Subsection.

\subsection{Results}
{\it SP-I phase.}-- For  a large tunneling  where
$\alpha^2+\beta^2>1$, the two layers are strongly coupled,  so that
the bilayer behaves like a single layer with  the 1D SOC oriented
along the diagonal axis.  In this case the single-particle
Hamiltonian (\ref{Heff})  yields a pair of degenerate  ground
eigenstates with  wave vectors $\pm\mathbf{Q}$. Across a critical
value of $g_{\uparrow\downarrow}/g$,  the condensate transits from
the PW  with a single wave vector to the normal SP-I phase, which is
characterized by the wave function  involving two wave-vectors
$\frac{1}{\sqrt{2}}\phi_{+\mathbf{Q}}e^{i\mathbf{Q}\cdot\mathbf{r}}
+\frac{1}{\sqrt{2}}\phi_{-\mathbf{Q}}e^{-i\mathbf{Q}\cdot\mathbf{r}}$.
In Fig. \ref{stripe-I} we see that, due to the nonvanishing
intra-layer coupling $\beta$,  the total density
$\rho_j(\mathbf{r})=|\psi_{j\uparrow}(\mathbf{r})|^2
+|\psi_{j\downarrow}(\mathbf{r})|^2$ in each layer  modulates for
the SP-I phase  \cite{Martone}.
\begin{figure}
\includegraphics[width= 0.47\textwidth]{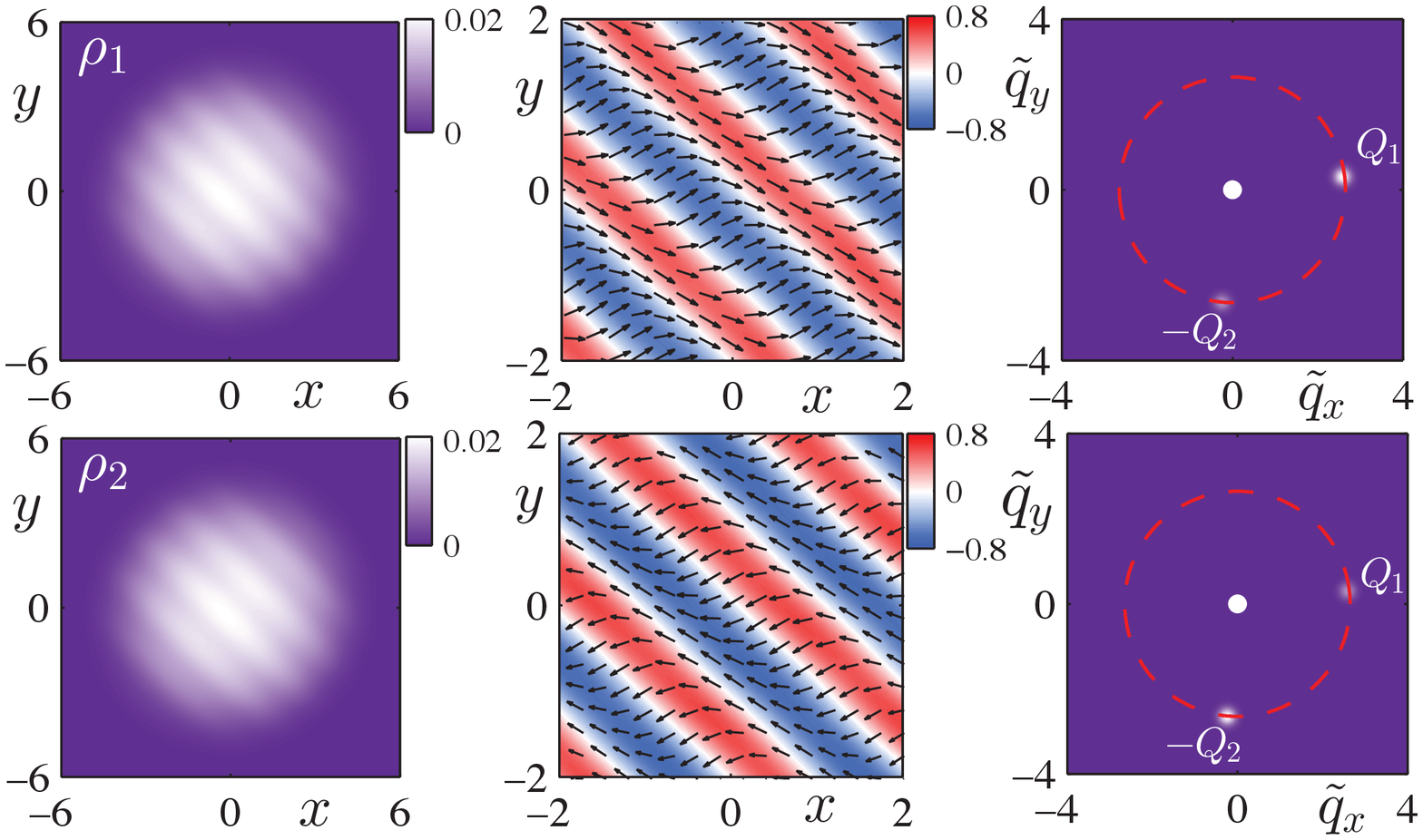}
\caption{(color online)   Total density profiles $\rho_j$ (left),
spin textures $\mathbf{S}_j$ (middle), and  the corresponding
momentum distributions (right) in the first (up) and second (bottom)
layer for the  SP-II phase with $\alpha=0.5$ and
$g_{\uparrow\downarrow}/g=0.9$. The color in the spin textures
indicates the magnitude of $S_{jz}$. \label{stripe}}
\end{figure}

One can define the spin texture for each layer.  For this purpose,
let us introduce a normalized two-component spinor
$\chi_{j}(\mathbf{r})=[\chi_{j\uparrow}(\mathbf{r}),
\chi_{j\downarrow}(\mathbf{r})]^T=[|\chi_{j\uparrow}|
e^{i\theta_{j\uparrow}},
|\chi_{j\downarrow}|e^{i\theta_{j\downarrow}}]^T$ and decompose the
wave function $\psi_{j}(\mathbf{r})$ as
$\psi_{j}(\mathbf{r})=\sqrt{\rho_{j}(\mathbf{r})}
\chi_{j}(\mathbf{r})$, where  $\chi_{j}$ satisfies
$|\chi_{j\uparrow}|^2+|\chi_{j\downarrow}|^2=1$ \cite{Kasamatsu}.
The spin texture can be represented by a vector
$\mathbf{S}_j=(2|\chi_{j\uparrow}||
\chi_{j\downarrow}|\cos(\theta_{j\uparrow}-\theta_{j\downarrow}),
-2|\chi_{j\uparrow}||
\chi_{j\downarrow}|\sin(\theta_{j\uparrow}-\theta_{j\downarrow}),
|\chi_{j\uparrow}|^2-|\chi_{j\downarrow}|^2)$.  It can be seen that
the density modulation are accompanied by the spin stripes with a
similar modulation, as depicted in Fig. \ref{stripe-I}.

{\it SP-II phase.}--Next, we discuss the parameter region
$\alpha^2+\beta^2<1$.  Here the inter-layer tunneling  mixes the
states belonging to different layers  in a more sophisticated  way,
so various nontrivial ground state configurations may appear. In
Fig. \ref{stripe} we show the numerical results of the total density
profiles, spin textures and  the corresponding momentum
distributions for a new type of  the stripe phase (SP-II). At first
sight, it appears that the density profile of SP-II is similar to
that of  the normal SP-I. However, when we turn to the momentum
space, the two types of the stripe phases differ dramatically. For
the SP-I, the momentum distribution in each layer comprises a pair
of opposite wave vectors $\pm\mathbf{Q}$  which conserve the TR
symmetry. Intriguingly, we find that although the ground state wave
function of SP-II remains a superposition of two wave vectors, the
comprising wave vectors are neither $\pm\mathbf{Q}_1$ nor
$\pm\mathbf{Q}_2$. Instead,  the SP-II becomes a superposition of
the waves with $\mathbf{Q}_1$ and $-\mathbf{Q}_2$, spontaneously
breaking the TR symmetry. In the momentum representation, the  SP-II
phase atoms are predominantly located at $\mathbf{Q}_1$ (in the
$\hat{x}$-direction)  in one layer, whereas  in another layer they
are concentrated at $-\mathbf{Q}_2$,  along the $\hat{y}$-direction.
Moreover, the resulting spin texture of SP-II phase exhibits a
chiral spin helix as shown in Fig. \ref{stripe}.
\begin{figure}
\includegraphics[width= 0.47\textwidth]{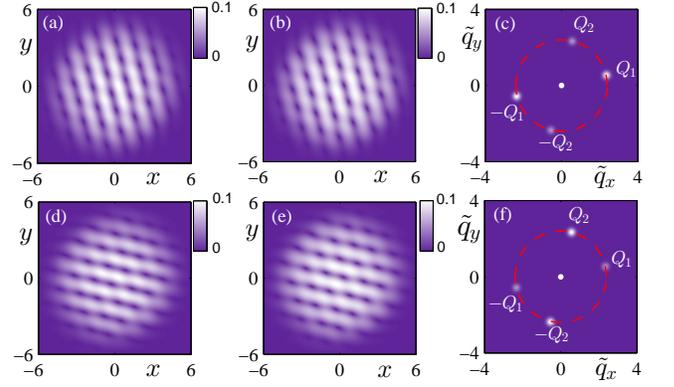}
\caption{(color online) Numerical density profiles of (a,d)
$|\psi_{j\uparrow}(\mathbf{r})|$ and (b,e)
$|\psi_{j\downarrow}(\mathbf{r})|$ for the FSL phase in each layer
($\alpha=0.8$, $g_{\uparrow\downarrow}/g=1.05$). (c,f)  The
corresponding momentum distributions of  first and second layer.
\label{vortex}}
\end{figure}

\begin{figure}
\includegraphics[width= 0.4\textwidth]{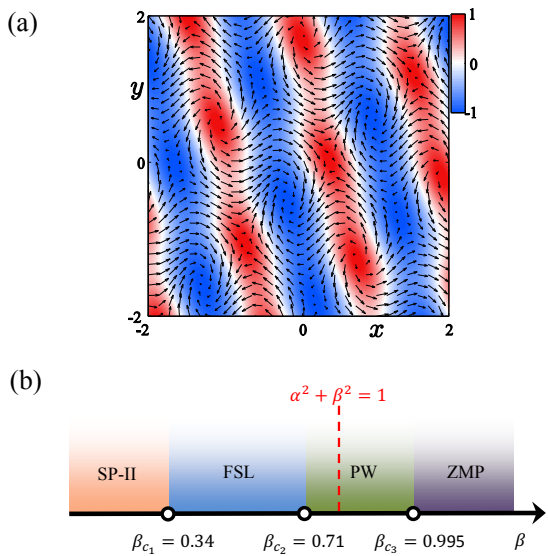}
\caption{(color online) (a) Spin texture of the up layer for the FSL
phase represented in Fig. \ref{vortex}.  The color in the spin
texture indicates the magnitude of $S_{1z}$. (b) Phase transitions
as a function of the intra-layer coupling $\beta$, where the
inter-layer tunneling is fixed by $\alpha=0.6$, and
$g_{\uparrow\downarrow}/g=0.9$. \label{texture}}
\end{figure}

{\it FSL phase.}--Beyond the SP-II, another  distinctive feature in
Fig. \ref{phase1} is that a fractionalized skyrmion lattice (FSL)
emerges in the ground state. In Fig. \ref{vortex}, a vortex lattice
structure can be seen in the density profiles of each spin
component. The lattices of both spin components interlace mutually,
forming a coreless structure in each layer. Most notably, the
momentum distributions display two pairs of TR invariant momenta, as
shown in Fig. \ref{vortex}(c,f). The atoms in each layer tend to be
mainly located at $\pm\mathbf{Q}_1$ and $\pm\mathbf{Q}_2$,
respectively, to make the energy favorable.  This indicates that the
underlying mechanism of the vortex lattices arises from the
four-wave interference with a nontrivial phase structure. Fig.
\ref{texture}(a) shows the spin texture of the  upper layer (the
lower layer  yields the analogous results), where  one can see
clearly a lattice of  skyrmions and antiskyrmions interlacing with
each other.

To further characterize this state, let us calculate the topological
charge  $Q_j=\int_{\rm unit\> cell} d^2 \mathbf{r} q_j(\mathbf{r})$
for $j$-th layer, where the topological density is given by
$q_j(\mathbf{r})=\frac{1}{8\pi}\epsilon^{\mu\nu}\mathbf{S}_j
\cdot\partial_\mu\mathbf{S}_j \times\partial_\nu\mathbf{S}_j$. Note
that, the limits of integration in the topological charge $Q_j$ are
defined over the unit cell of the lattice.  However, since the
boundary between a skyrmion and antiskyrmion is hard to be
explicitly  discriminated, the integral  only approximately equals
to a half integer.  In practice, one may integrate over the whole
area of the system and find that the total topological charge
$Q_j^T=\int_{\rm whole} d^2 \mathbf{r} q_j(\mathbf{r})$ vanishes. On
the other hand, we compute the integral of the absolute value of the
topological density ${Q_j^T}^\prime=\int_{\rm whole} d^2 \mathbf{r}
|q_j(\mathbf{r})|$. This yields an integer $I_j$. Then, by counting
the total number $N_j$ of the topological defects, we obtain the
topological charge of the interlacing skyrmion and antiskyrmion,
$Q_j=\pm I_j/N_j=\pm1/2$ \cite{Su,Guo1}. This confirms the formation
of a FSL, an intriguing topological ground state emerging in such a
homogeneous system. It should be noted that the FSL phase cannot
exist along the $\alpha=0$ line in the phase diagram of Fig.
\ref{phase1}.  In that case  the two layers are decoupled, each of
them having two degenerate energy minima in different ($\hat{x}$ or
$\hat{y}$) directions.

{\it Phase diagram and TP.}--When the inter-layer tunneling is tuned
to $\alpha^2+\beta^2=1$, a TP may occur on the critical line marked
by  a circle in Fig. \ref{phase1}. Indeed,  starting from the SP-I
phase and decreasing $\alpha$, the system would first transit across
the critical line to  the SP-II before entering into the FSL phase.
On the other hand,  the PW phase extends into the region below the
critical line and transits to  the FSL directly. Therefore,  the TP
occurs among the four different phases.  This is can also be clearly
demonstrated in the variational phase diagram.

Having studied the $\alpha - g_{\uparrow\downarrow}/g$ phase
diagram, we  next discuss the effects of  the intra-layer coupling
$\beta$ which can be varied conveniently in experiments. For this
purpose, we take the parameters $\alpha=0.6$ and
$g_{\uparrow\downarrow}/g=0.9$  as an illustration. In Fig.
\ref{texture}(b), we show that, with increasing  $\beta$, the system
first undergoes a transition from  the SP-II to FSL phase at a
critical point $\beta_{c_1}$. Subsequently the system enters into
the PW phase near the critical line. Finally, as  the intra-layer
coupling approaches $\beta_{c_3}\simeq1$, the momenta of  the energy
minima shrink to $\mathbf{Q}=\mathbf{0}$ and the atoms condense in
the zero-momentum phase (ZMP). All these  phases can be observed
through the spin-resolved time-of-flight measurements of the density
profiles, the momentum distributions, and spin textures.

\section{Discussion and Conclusion}
Finally, we discuss the experiment related issues. The result of our
paper can be applied to a number of systems involving two atomic
internal  states coupled by laser beams with the recoil, such as two
magnetic sub-levels of the $F=1$ ground state manifold of the
$^{87}$Rb-type alkali atoms  \cite{Lin}  or the spin-singlet ground
state and a long-lived spin triplet excited state of the
alkaline-earth atoms \cite{Gerbier}. Here we consider the former
example. We take $N=10^{4}$ $^{87}$Rb atoms with the trapping
frequencies $\left( \omega _{\bot },\omega _{z}\right) = 2\pi \times
\left(10,400\right)$ Hz. For the wave length of Raman lasers
$\lambda _{L}\simeq 804.1$ nm \cite{Lin}, we have $E_{\kappa}\simeq
11\hbar \omega _{\bot }$. The scattering lengths for two the spin
states $|F=1,m_F=0\rangle\equiv|\uparrow\rangle$ and
$|F=1,m_F=-1\rangle\equiv|\downarrow\rangle$,  used in ref.
\cite{Lin}, are usually parameterized as  \cite{Ho1,Ohmi}
$a_\uparrow=c_0$ and $a_\downarrow=a_{\uparrow\downarrow}=c_0+c_2$,
with $c_0=7.79\times 10^{-12}$ Hz cm$^3$ and $c_2=-3.61\times
10^{-14}$ Hz cm$^3$. The corresponding intra- and inter-species
atomic interactions are given by
$g_{\uparrow,\downarrow}=\sqrt{2\pi}Na_{\uparrow,\downarrow}/\xi_z$
and
$g_{\uparrow\downarrow}=\sqrt{2\pi}Na_{\uparrow\downarrow}/\xi_z$,
with $\xi_z=\sqrt{\hbar/m\omega_z}$. We would like to point out that all the parameters we choose are limited to a weakly interacting region, in which the coherence length is large in comparison with the size of the trap $\xi_z$ so that the mean-field analysis is applicable.

\begin{figure}
\includegraphics[width=0.45\textwidth]{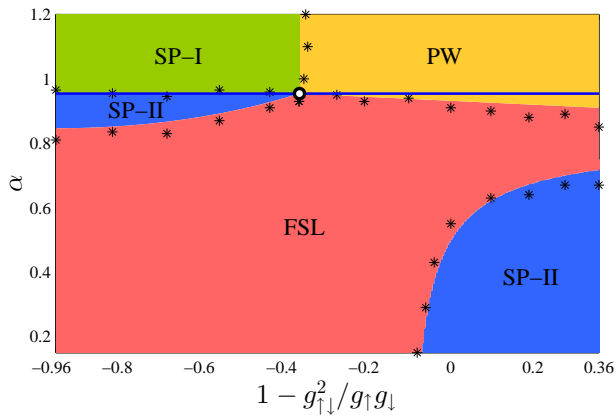}
\caption{(color online) Phase diagram  as a function of the
dimensionless inter-layer tunneling $\alpha$ and
$1-g_{\uparrow\downarrow}^2/g_{\uparrow}g_{\downarrow}$ for the
asymmetric intra-species interaction
$g_{\uparrow}/g_{\downarrow}=0.95$. The dimensionless intra-layer
coupling is set to be $\beta=0.3$. The stars represent the phase
boundaries determined from the numerical simulations and the colored
regions are determined by the variational results. The horizontal
solid line marks the critical line $\alpha^{2}+\beta^{2}=1$.
\label{phase2}}
\end{figure}

Note that, the intra-species interaction is nearly symmetric with
$g_\uparrow/g_\downarrow=1.0047$, so the phase diagram of Fig.
\ref{phase1} can be applied directly. However, it is important to
discuss a more general case with asymmetric intra-species
interaction $g_\uparrow\neq g_\downarrow$.  To check whether the
predicted new  phases and TP  are preserved in the asymmetric case,
we take $g_{\uparrow}/g_{\downarrow}=0.95$ as an example, and
calculate the phase diagram shown in Fig. \ref{phase2}. We find that
although the phase boundaries get modified, the phase diagram as a
function of $\alpha$ and
$1-g_{\uparrow\downarrow}^2/g_{\uparrow}g_{\downarrow}$  bears
similar structure as that of the  symmetric situation, demonstrating
that these are the unique and universal features of the  bilayer
system for a wide range of atomic interaction parameters. Note that
the TP still appears on the critical line, but gets shifted by the
asymmetric intra-species interaction.

In summary, we  have proposed a tunneling-assisted SOC in bilayer
BECs. This scheme can be realized in a straightforward manner by
coupling the individual Raman transition induced 1D SOC through the
inter-layer laser-assisted tunneling. Due to  an interplay between
the inter-layer tunneling, intra-layer  SOC and atomic interactions,
the ground states display a diverse phase diagram.  It is
demonstrated that a new type of stripe phase which breaks the TR
symmetry and a fractionalized skyrmion lattice emerge spontaneously
in the ground states. Significantly, we predict the occurrence of  a
characteristic tetracritical point, where the four different phases
merge together.  Such distinctive features are within the reach of
current experiments with ultracold atoms.

\section*{ACKNOWLEDGMENTS} We acknowledge H. Zhai, X. F. Zhang,
S.-C. Gou, and  L. Santos for helpful discussions. This work is
supported by NCET, NSFC under grants Nos. 11404225, 11474205,
NKBRSFC under grants Nos. 2011CB921502, 2012CB821305, and the
European Social Fund under the Global Grant measure.

\setcounter{equation}{0}
\renewcommand{\theequation}{A\arabic{equation}}

\section*{APPENDIX A: GROUND STATE MANIFOLD OF THE SINGLE-PARTICLE HAMILTONIAN}

In this section, we shall obtain the ground eigenstates and the
corresponding eigen energies of the single-particle Hamiltonian
described by Eq. (\ref{Heff}). To make the Hamiltonian more compact,
the second-quantized Hamiltonian (\ref{Heff}) may be expressed in
terms of a four-component  column spinor
$\hat{\Psi}(\mathbf{r})=[\hat{\psi}_{1\uparrow}(\mathbf{r}),
\hat{\psi}_{1\downarrow}(\mathbf{r}),\hat{\psi}_{2\uparrow}(\mathbf{r}),
\hat{\psi}_{2\downarrow}(\mathbf{r})]^{T}$ containing operators
which annihilate an atom in a specific layer $j=1,2$ and a specific
spin state $\gamma=\uparrow,\downarrow$
\begin{equation}
\hat{H}_{{\rm eff}}=\int
d^{2}\mathbf{r}\hat{\Psi}^{\dag}(\mathbf{r})H_{\rm
eff}\hat{\Psi}(\mathbf{r})\,,\label{four-spinor}
\end{equation}
where $H_{\rm eff}$ is the $4\times4$ matrix Hamiltonian
\begin{equation}
H_{\rm eff}=\frac{q^{2}}{2}I_{4}+\left(\begin{array}{cccc}
\Omega & \kappa q_{x} & J & 0\\
\kappa q_{x} & -\Omega & 0 & J\\
J & 0 & \Omega & \kappa q_{y}\\
0 & J & \kappa q_{y} & -\Omega
\end{array}\right)\,,\label{eq:H_0-initial}
\end{equation}
and
$\hat{\Psi}^{\dagger}(\mathbf{r})=[\hat{\psi}_{1\uparrow}^{\dagger}(\mathbf{r}),
\hat{\psi}_{1\downarrow}^{\dagger}(\mathbf{r}),\hat{\psi}_{2\uparrow}^{\dagger}(\mathbf{r}),
\hat{\psi}_{2\downarrow}^{\dagger}(\mathbf{r})]$ is the Hermitically
conjugated row spinor.

To determine the eigen energies and the corresponding eigenstates of
the single-particle  problem, we shall  analyze the latter matrix
Hamiltonian $H_{\rm eff}$. In a homogeneous system, the momentum is
a conserved quantity, so the eigen functions of $H_{\rm eff}$ are
the four-component plane waves
$\psi_{\mathbf{\mathbf{\widetilde{q}}}}(r)=[\psi_{1\uparrow}(\mathbf{\widetilde{q}}),
\psi_{1\downarrow}(\mathbf{\widetilde{q}}),\psi_{2\uparrow}(\mathbf{\widetilde{q}}),
\psi_{2\downarrow}(\mathbf{\widetilde{q}})]^{T}e^{i\mathbf{\widetilde{q}}\cdot\mathbf{r}}
\equiv\phi_{\mathbf{\widetilde{q}}}e^{i\mathbf{\widetilde{q}\cdot
r}}$. Here $\widetilde{\mathbf{q}}=\mathbf{q}/\kappa$ is a
dimensionless momentum, and
$\psi_{j\gamma}(\mathbf{\widetilde{q}})e^{i\mathbf{\widetilde{q}\cdot
r}}$ represents the probability amplitude to find the atom in the
$j$-th layer ($j=1,2$) and the internal state
$\gamma=\uparrow,\downarrow$.

The eigen equation reads:
\begin{eqnarray}
H_{\rm
eff}\phi_{\mathbf{\widetilde{q}}}=E\phi_{\mathbf{\widetilde{q}}}\,.\label{Eign-eq}
\end{eqnarray}
It is convenient to rewrite the $4\times4$ matrix Hamiltonian
$H_{\rm eff}$ in terms of a $2\times2$ matrix with elements
containing the unit matrix $I_{2}$ and the Pauli matrices $\sigma_x$
and $\sigma_z$:
\begin{eqnarray}
H_{\rm eff}=E_{\kappa}\left(\begin{array}{cc}
\beta\sigma_{z}+2\tilde{q}_{x}\sigma_{x} & \alpha I_{2} \\
\alpha I_{2} & \beta\sigma_{z}+2\tilde{q}_{y}\sigma_{x}\\
\end{array}\right)\,,\label{eq:H_0-compact}
\end{eqnarray}
where $\beta=\Omega/E_{\kappa}$, $\alpha=J/E_{\kappa}$ are,
respectively, the dimensionless energies of the intra-layer coupling
and inter-layer tunneling measured in the units of the recoil energy
$E_{\kappa}=\kappa^{2}/2$. In Eq. (\ref{eq:H_0-compact}) we have
omitted the overall energy shift  $q^{2}/2$ which is to be
substracted from  the eigen-energy $E$ in Eq. (\ref{Eign-eq}).

Combining Eqs. (\ref{Eign-eq})  and (\ref{eq:H_0-compact}), the
dimensionless eigen energy $\tilde{\omega}=(E-q^2/2)/E_{\kappa}$
satisfies the  equation:
\begin{eqnarray}
\left|\begin{array}{cc}
\beta\sigma_{z}+2\tilde{q}_{x}\sigma_{x}-\tilde{\omega} I_{2} & \alpha I_{2} \\
\alpha I_{2} &\beta\sigma_{z}+2\tilde{q}_{y}\sigma_{x} - \tilde{\omega }I_{2}\\
\end{array}\right|=0\,.
\label{matrix222}
\end{eqnarray}
By using the block matrix theory \cite{Eves}, we can rewrite Eq.
(\ref{matrix222}) as
\begin{equation}
\left|\left(\beta\sigma_{z}+2\tilde{q}_{x}\sigma_{x}-\tilde{\omega}
I_{2} \right)
\left(\beta\sigma_{z}+2\tilde{q}_{y}\sigma_{x}-\tilde{\omega} I_{2}
\right) -\alpha^{2} I_{2}\right|=0\,,\label{matrix22-Pauli-form}
\end{equation}
and hence
\begin{equation}
\left| g I_{2}-2\tilde{\omega}\left(\tilde{q}_{x}+\tilde{q}_{y}\right)\sigma_{x}
-2\tilde{\omega}\beta\sigma_{z}+i2\beta\left(\tilde{q}_{x}-\tilde{q}_{y}\right)\sigma_{y}\right|=0\,,
\label{matrix22-Pauli-form-1}
\end{equation}
with
\[g=\tilde{\omega}^{2}+\beta^{2}+4\tilde{q}_{x}\tilde{q}_{y}-\alpha^{2}\,.
\]
 This yields the following eigenvalue equation:
\begin{equation}
g^{2}=4\tilde{\omega}^{2}\left(\tilde{q}_{x}+\tilde{q}_{y}\right)^{2}
+4\tilde{\omega}^{2}\beta^{2}-4\beta^{2}\left(\tilde{q}_{x}-\tilde{q}_{y}\right)^{2}\,.
\label{Eigenvalue-eq}
\end{equation}
After direct calculations one arrives at  a bi-quadratic equation
\begin{equation}
\left[\tilde{\omega}^2-(2\widetilde{q}^{2}
+\alpha^{2}+\beta^{2})\right]^2=A^2\,, \label{Eigenvalue-eq1}
\end{equation}
providing  four branches of energy spectra
\begin{equation}
E_{\pm,\pm}(\mathbf{\widetilde{q}})/E_{\kappa}=\widetilde{q}^{2}\pm\sqrt{2\widetilde{q}^{2}
+\alpha^{2}+\beta^{2}\pm A}\,,\label{eigenenergies}
\end{equation}
where
\begin{equation}
A=2\sqrt{(\widetilde{q}_{x}+\widetilde{q}_{y})^{2}
[(\widetilde{q}_{x}-\widetilde{q}_{y})^{2}+\alpha^{2}]+\alpha^{2}\beta^{2}}\,.
\label{A-expression}
\end{equation}

In what follows, we  will focus on the lowest branch
$E_{-,+}(\mathbf{\widetilde{q}})$,  and determine the energy minima
which play an important role in   formation of the ground state
configurations.  For this, one needs to identify the points where
$\partial
E_{-,+}(\mathbf{\widetilde{q}})/\partial\widetilde{q}_{x}=0$ and
$\partial
E_{-,+}(\mathbf{\widetilde{q}})/\partial\widetilde{q}_{y}=0$, giving
\begin{eqnarray}
 2\widetilde{q}_{x}B-2\widetilde{q}_{x}
 -\frac{2\widetilde{q}_{x}(\widetilde{q}_{x}^{2}-\widetilde{q}_{y}^{2})
 +\alpha^{2}(\widetilde{q}_{x}+\widetilde{q}_{y})}{\sqrt{(\widetilde{q}_{x}
 +\widetilde{q}_{y})^{2}[(\widetilde{q}_{x}-\widetilde{q}_{y})^{2}+\alpha^{2}]
 +\alpha^{2}\beta^{2}}}\!\!&=&\!\!0\,,\nonumber \\
 2\widetilde{q}_{y}B-2\widetilde{q}_{y}
 -\frac{2\widetilde{q}_{y}(\widetilde{q}_{y}^{2}-\widetilde{q}_{x}^{2})
 +\alpha^{2}(\widetilde{q}_{x}+\widetilde{q}_{y})}{\sqrt{(\widetilde{q}_{x}
 +\widetilde{q}_{y})^{2}[(\widetilde{q}_{x}-\widetilde{q}_{y})^{2}+\alpha^{2}]
 +\alpha^{2}\beta^{2}}}\!\!&=&\!\!0\,,\nonumber \\
\label{eqs}
\end{eqnarray}
with $B\equiv\sqrt{2\widetilde{q}^{2}+\alpha^{2}+\beta^{2}
 +A}$. For the most interesting case where $\alpha^{2}+\beta^{2}<1$,
 the above two equations yield four  chiral states with minimum
energies  at
$\pm\mathbf{Q}_{1}=\pm(\widetilde{q}_{0}^{+},\widetilde{q}_{0}^{-})$
and
$\pm\mathbf{Q}_{2}=\pm(\widetilde{q}_{0}^{-},\widetilde{q}_{0}^{+})$.
 Here $\widetilde{q}_{0}^{\pm}=\frac{1}{2}(\sqrt{Q_{0}^{2}
+\alpha^{2}/2}\pm\sqrt{Q_{0}^{2}-\alpha^{2}/2})$, and
$Q_{0}=|\mathbf{Q}_{1,2}|$  satisfy a nonlinear equation
\begin{eqnarray}
\sqrt{2Q_{0}^{2}+\alpha^{2}+\beta^{2}+C}
-\frac{Q_{0}^{2}+\alpha^{2}/2}{\sqrt{(Q_{0}^{2}
+\alpha^{2}/2)^{2}+\alpha^{2}\beta^{2}}}&=&1\,,\nonumber\\
\label{nonlinear}
\end{eqnarray}
with $C\equiv2\sqrt{(Q_{0}^{2}+\alpha^{2}/2)^{2}
+\alpha^{2}\beta^{2}}$.

The corresponding  eigen function  is given by, for four degenerate
energy minima at $\mathbf{\widetilde{q}}=\pm\mathbf{Q}_1$ and
$\mathbf{\widetilde{q}}=\pm\mathbf{Q}_2$
\begin{eqnarray}
\psi_{\mathbf{\widetilde{q}}}\!\!=\!\!f(\mathbf{\widetilde{q}})
\!\left(\begin{array}{c}
\alpha\left[\beta\xi-\beta^{2}-\zeta-\left(\widetilde{q}_{x}
+\widetilde{q}_{y}\right)^{2}\right]\\
\alpha\left[\beta\left(\widetilde{q}_{x}-\widetilde{q}_{y}\right)
-\xi\left(\widetilde{q}_{x}+\widetilde{q}_{y}\right)\right]\\
\left(\beta-\xi\right)\left(\widetilde{q}_{x}^{2}-\widetilde{q}_{y}^{2}
-\zeta\right)-\alpha^{2}\beta\\
2\widetilde{q}_{y}\left(\zeta-\widetilde{q}_{x}^{2}+\widetilde{q}_{y}^{2}\right)
+\alpha^{2}\left(\widetilde{q}_{x}+\widetilde{q}_{y}\right)
\end{array}\right)\!e^{i\mathbf{\widetilde{q}\cdot r}}\,,\nonumber\\
\label{eigen-wave}
\end{eqnarray}
where
$\zeta=\sqrt{\alpha^{2}\beta^{2}+\alpha^{2}\left(\widetilde{q}_{x}
+\widetilde{q}_{y}\right)^{2}+\left(\widetilde{q}_{x}^{2}
-\widetilde{q}_{y}^{2}\right)^{2}}$,
$\xi=\sqrt{\alpha^{2}+\beta^{2}+2\widetilde{q}^{2}+2\zeta}$, and
 $f(\mathbf{\widetilde{q}})$ is the normalized coefficient.

\section*{APPENDIX B: ENERGY FUNCTIONAL FOR VARIATIONAL ANSATZ}
\setcounter{equation}{0}
\renewcommand{\theequation}{B\arabic{equation}}
To calculate the mean-field energy under the variational Ansatz,
it's convenient to rewrite the interacting Hamiltonian (\ref{intH})
in the four-spinor representation $\hat{\Psi}$, which is given by
\begin{eqnarray}
\hat{H}_{\rm int}=\frac{1}{2}\int d^2
\mathbf{r}\sum^6_{m=1}b_m\left(\hat{\Psi}^\dag\mathcal{M}_m\hat{\Psi}\right)^2\,.\label{intH1}
\end{eqnarray}
Here $\mathcal{M}_m$  are the four  $4\times4$ matrices  which can
be represented as: $\mathcal{M}_1=I_4$,
$\mathcal{M}_2=\left(\begin{array}{cc} I_2
& 0 \\
0
& -I_2\\
\end{array}\right)$, $\mathcal{M}_3=\left(\begin{array}{cc} \sigma_z
& 0 \\
0
& \sigma_z\\
\end{array}\right)$, $\mathcal{M}_4=\left(\begin{array}{cc} \sigma_z
& 0 \\
0
& -\sigma_z\\
\end{array}\right)$,
$\mathcal{M}_5=\frac{1}{2}(\mathcal{M}_1+\mathcal{M}_3)$, and
$\mathcal{M}_6=\frac{1}{2}(\mathcal{M}_2+\mathcal{M}_4)$; $b_m$ are
the the coefficients with
$b_1=b_2=(g_\uparrow+g_{\uparrow\downarrow})/4$,
$b_3=b_4=(g_\uparrow-g_{\uparrow\downarrow})/4$, and
$b_5=b_6=(g_\downarrow-g_\uparrow)/2$.

We take the Ansatz
$\psi_G\equiv\langle\hat{\Psi}\rangle=\sum_{j=1,2;\pm}a_{j\pm}\psi_{\pm
\mathbf{Q}_j}$, where
$\psi_{\pm\mathbf{Q}_{1,2}}\equiv\phi_{\pm\mathbf{Q}_{1,2}} e^{\pm
i\mathbf{Q}_{1,2}\cdot\mathbf{r}}$ denote four eigen functions which
correspond to the minimum energy  and are given by Eq.
(\ref{eigen-wave}). The complex  amplitudes $a_{j\pm}$  satisfy the
normalization condition $\sum_{j,\pm}|a_{j\pm}|^2=1$. Subsequently,
by replacing $\hat{\Psi}$ in Eq. (\ref{intH1}) with $\psi_G$, we
derive the mean-field interacting energy functional
$\mathcal{E}[a_{1\pm},a_{2\pm}]=\langle\hat{H}_{\rm{int}}\rangle$ as
shown in Eq. (\ref{variation}). The corresponding coefficients
$C_{1-5}$ in Eq. (\ref{variation}) read
\begin{eqnarray}
C_1&=&\frac{1}{2}\sum_mb_m\left(\bar{\phi}_{\mathbf{Q}_1}
\mathcal{M}_m\phi_{\mathbf{Q}_1}\right)^2\,,\nonumber\\
C_2&=&\frac{1}{2}\sum_mb_m\left[\left(\bar{\phi}_{\mathbf{Q}_1}
\mathcal{M}_m\phi_{\mathbf{Q}_1}\right)\left(\bar{\phi}_{\mathbf{Q}_2}
\mathcal{M}_m\phi_{\mathbf{Q}_2}\right)\right.\nonumber\\
&&\left.+\left(\bar{\phi}_{\mathbf{Q}_1}
\mathcal{M}_m\phi_{\mathbf{Q}_2}\right)\left(\bar{\phi}_{\mathbf{Q}_2}
\mathcal{M}_m\phi_{\mathbf{Q}_1}\right)\right]\,,\nonumber\\
C_3&=&\frac{1}{2}\sum_mb_m\left[\left(\bar{\phi}_{\mathbf{Q}_1}\mathcal{M}_m
\phi_{\mathbf{Q}_1}\right)\left(\bar{\phi}_{-\mathbf{Q}_2}\mathcal{M}_m
\phi_{-\mathbf{Q}_2}\right)\right.\nonumber\\
&&\left.+\left(\bar{\phi}_{\mathbf{Q}_1}\mathcal{M}_m
\phi_{-\mathbf{Q}_2}\right)\left(\bar{\phi}_{-\mathbf{Q}_2}\mathcal{M}_m
\phi_{\mathbf{Q}_1}\right)\right]\,,\nonumber\\
C_4&=&\frac{1}{2}\sum_mb_m\left[\left(\bar{\phi}_{\mathbf{Q}_1}
\mathcal{M}_m\phi_{\mathbf{Q}_1}\right)\left(\bar{\phi}_{-\mathbf{Q}_1}
\mathcal{M}_m\phi_{-\mathbf{Q}_1}\right)\right.\nonumber\\
&&\left.+\left(\bar{\phi}_{\mathbf{Q}_1}
\mathcal{M}_m\phi_{-\mathbf{Q}_1}\right)\left(\bar{\phi}_{-\mathbf{Q}_1}
\mathcal{M}_m\phi_{\mathbf{Q}_1}\right)\right]\,,\nonumber\\
C_5&=&\frac{1}{2}\sum_mb_m\left[\left(\bar{\phi}_{\mathbf{Q}_1}
\mathcal{M}_m\phi_{\mathbf{Q}_2}\right)\left(\bar{\phi}_{-\mathbf{Q}_1}
\mathcal{M}_m\phi_{-\mathbf{Q}_2}\right)\right.\nonumber\\
&&\left.+\left(\bar{\phi}_{\mathbf{Q}_1}
\mathcal{M}_m\phi_{-\mathbf{Q}_2}\right)\left(\bar{\phi}_{-\mathbf{Q}_1}
\mathcal{M}_m\phi_{\mathbf{Q}_2}\right)+\rm
H.c.\right]\,.\nonumber\\
\label{coefficient2}
\end{eqnarray}
The variational phase diagram is obtained by minimizing the energy
$\mathcal{E}[a_{1\pm},a_{2\pm}]$.

\end{document}